\documentstyle[preprint,tighten,aps,epsfig]{revtex}
\begin{document}

\title{$\Lambda NN$ and $\Sigma NN$ systems at threshold: II. The effect
of $D$ waves}

\author{H. Garcilazo$^{(1)}$, A. Valcarce$^{(2)}$, and 
T. Fern\'andez-Caram\'es$^{(3)}$} 
\address{$(1)$ Escuela Superior de F\'\i sica y Matem\'aticas, 
Instituto Polit\'ecnico Nacional, 
Edificio 9, 07738 M\'exico D.F., Mexico}
\address{$(2)$ Departamento de F\'\i sica Fundamental, 
Universidad de Salamanca, E-37008 Salamanca, Spain}
\address{$(3)$ Departamento de F\' \i sica Te\'orica e IFIC, 
Universidad de Valencia - CSIC, 
E-46100 Burjassot, Valencia, Spain}
\maketitle

\date{\today}
\begin{abstract}
Using the two-body interactions obtained from a chiral constituent 
quark model we study all $\Lambda NN$ and $\Sigma NN$ states with
$I=0,1,2$ and $J=1/2,3/2$ at threshold, taking into account all
three-body configurations with $S$ and $D$ wave components. We 
constrain further the limits 
for the $\Lambda N$ spin-triplet scattering 
length $a_{1/2,1}$. Using the
hypertriton binding energy we find a narrow interval
for the possible values of the 
$\Lambda N$ spin-singlet scattering 
length $a_{1/2,0}$. We found that the
$\Sigma NN$ system has a quasibound state in the $(I,J) = (1,1/2)$
channel very near threshold with a width of about 2.1 MeV.
\end{abstract}

\maketitle

\section{Introduction}

The chiral constituent quark model has been very successful in the
simultaneous description of the baryon-baryon interaction and the baryon 
spectrum as well as in the study of the two- and
three-baryon bound-state problem for the nonstrange sector \cite{Val05}.
A simple generalization of this model to the strange sector has been
applied to study the meson and baryon spectra \cite{Gar05} and the 
$\Sigma NN$ bound-state problem \cite{Ter06}.
Recently, a more elaborated description of the model 
was developed in Ref. \cite{GFV07},
where the $\Lambda NN$ system was also studied.

In Ref. \cite{GFV07} we studied the $\Lambda NN$ and $\Sigma NN$ systems
at threshold by solving the Faddeev equations of the coupled 
$\Lambda NN - \Sigma NN$ system in the case of pure $S$ wave configurations
for the channels $(I,J)$ with $I=0,1,2$ and $J=1/2,3/2$.
However, since the hyperon-nucleon and nucleon-nucleon interactions 
contain sizeable tensor terms there is a coupling between the $\ell=0$
and $\ell=2$ baryon-baryon channels and between the hyperon-nucleon-nucleon
channels with $\ell=0$ and $\lambda=0$ to the channels with $\ell=2$ and
$\lambda=2$. 
The importance of the tensor force at the two-body level manifests itself 
dramatically in the case of the $\Sigma^- p\to\Lambda n$ process which is
dominated by the $\Sigma N(\ell=0)\to \Lambda N(\ell=2)$ transition
such that if one includes only the 
$\Sigma N(\ell=0)\to \Lambda N(\ell=0)$ transition it is practically
impossible to describe the cross section \cite{Ter06} (this problem
was first observed in Ref. \cite{Str95}). Thus, one expects that also 
at the three-body level the effect of the $D$ waves will be important.

In Refs. \cite{Ter06,GFV07} we considered all configurations
where the baryon-baryon subsystems are in an $S$ wave and the third particle
is also in an $S$ wave with respect to the pair. However, to construct the 
two-body $t-$matrices that serve as input of the Faddeev equations we
considered the full interaction including the contribution of the $D$ waves
and of course the coupling between the $\Sigma N$ and $\Lambda N$
subsystems (which is known as the truncated $t-$matrix approximation
\cite{BERTH}).
In Ref. \cite{GFV07} we found that our model with only $S$ waves is able 
to predict correctly the binding energy of the hypertriton, which is a bound
state in the channel $(I,J)=(0,1/2)$. We also found that the channel 
$(I,J)=(0,3/2)$ will develop a bound state if the triplet $\Lambda N$
scattering length $a_{1/2,1}$ is larger than 1.68 fm. 
In the case of the $\Sigma NN$ system the channel 
$(I,J)=(1,1/2)$ develops a quasibound state in some cases while the channel
$(I,J)=(0,1/2)$ is also attractive but unbound. 

In this work, we will
further pursue the study of the $\Lambda NN -\Sigma NN$ system at threshold
when the three-body $D$ wave components are considered. We will analyze
their effects comparing our results with those obtained when using
only three-body $S$ wave contributions. 
The structure of the paper is the following.
In the next section we will resume the basic aspects of 
the two-body interactions and we will present the
generalization of the Faddeev equations of Ref. \cite{GFV07} for
arbitrary orbital angular momenta. In section III we present our results
as compared to those of Ref. \cite{GFV07} to discuss the effect
of the three-body $D$ waves. Finally, in section IV we summarize our main
conclusions.

\section{Formalism}

\subsection{The two-body interactions}

The baryon-baryon interactions involved in the study of the coupled
$\Sigma NN - \Lambda NN$ system are obtained from the chiral 
constituent quark model \cite{Val05,Gar05}. In this model baryons
are described as clusters of three interacting massive (constituent) quarks,
the mass coming from the spontaneous breaking of chiral symmetry. The
first ingredient of the quark-quark interaction is a confining
potential ($CON$). Perturbative aspects of QCD are taken into account
by means of a one-gluon potential ($OGE$). Spontaneous breaking of 
chiral symmetry gives rise to boson exchanges
between quarks. In particular, there appear pseudoscalar 
boson exchanges and their corresponding scalar partners \cite{GFV07}. Thus,
the quark-quark interaction will read: 
\begin{equation}
V_{qq}(\vec{r}_{ij})=V_{CON}(\vec{r}_{ij})+V_{OGE}(\vec{r}_{ij})+V_{\chi}
(\vec{r}_{ij})+V_{S}(\vec{r}_{ij}) \,\,,
\label{int}
\end{equation}%
where the $i$ and $j$ indices are associated with $i$ and $j$ quarks
respectively, and ${\vec{r}}_{ij}$ stands for the interquark distance. 
$V_{\chi}$ denotes the pseudoscalar meson-exchange interaction
discussed in Ref. \cite{Ter06}, and $V_S$ stands for the scalar
meson-exchange potential described in Ref. \cite{GFV07}.
Explicit expressions of all the
interacting potentials and a more detailed discussion 
of the model can be found in Refs. \cite{Gar05,GFV07}.
In order to derive the local $B_1B_2\to B_3B_4$ potentials from the
basic $qq$ interaction defined above we use a Born-Oppenheimer
approximation. Explicitly, the potential is calculated as follows,

\begin{equation}
V_{B_1B_2 (L \, S \, T) \rightarrow B_3B_4 (L^{\prime}\, S^{\prime}\, T)} (R) =
\xi_{L \,S \, T}^{L^{\prime}\, S^{\prime}\, T} (R) \, - \, \xi_{L \,S \,
T}^{L^{\prime}\, S^{\prime}\, T} (\infty) \, ,  \label{Poten1}
\end{equation}

\noindent where

\begin{equation}
\xi_{L \, S \, T}^{L^{\prime}\, S^{\prime}\, T} (R) \, = \, {\frac{{\left
\langle \Psi_{B_3B_4 }^{L^{\prime}\, S^{\prime}\, T} ({\vec R}) \mid
\sum_{i<j=1}^{6} V_{qq}({\vec r}_{ij}) \mid \Psi_{B_1B_2 }^{L \, S \, T} ({\vec R%
}) \right \rangle} }{{\sqrt{\left \langle \Psi_{B_3B_4 }^{L^{\prime}\,
S^{\prime}\, T} ({\vec R}) \mid \Psi_{B_3B_4 }^{L^{\prime}\, S^{\prime}\, T} ({%
\vec R}) \right \rangle} \sqrt{\left \langle \Psi_{B_1B_2}^{L \, S \, T} ({\vec %
R}) \mid \Psi_{B_1B_2}^{L \, S \, T} ({\vec R}) \right \rangle}}}} \, .
\label{Poten2}
\end{equation}
In the last expression the quark coordinates are integrated out keeping $R$
fixed, the resulting interaction being a function of the $B_i-B_j$ 
relative distance. The wave function 
$\Psi_{B_iB_j}^{L \, S \, T}({\vec R})$ for the two-baryon system
is discussed in detail in Ref. \cite{Val05}.

\subsection{Faddeev equations at threshold}

Our method \cite{Ter06} to transform the Faddeev equations from being integral 
equations in two continuous variables into integral equations in just one
continuous variable is based in the expansion of the two-body $t-$matrices 
\begin{equation}
t_{i}(p_i,p^\prime_i;e)=\sum_{nr}P_n(x_i)
\tau_{i}^{nr}(e)P_r(x^\prime_i),
\label{for11}
\end{equation}
where $P_n$ and $P_r$ are Legendre polynomials,

\begin{equation}
x_{i}={\frac{p_{i}-b}{p_{i}+b}},  \label{for9}
\end{equation}%

\begin{equation}
x_{i}^\prime={\frac{p_{i}^\prime-b}{p_{i}^\prime+b}},  \label{for9p}
\end{equation}%
and $p_i$ and $p_i^\prime$ are the initial and final relative momenta of
the pair $jk$ while $b$ is a scale parameter on which the results do
not depend.

In Ref. \cite{GFV07} we wrote down the integral equations for $\beta d$
scattering at threshold with $\beta=\Sigma$ or $\Lambda$ 
including the full coupling between $\Lambda NN$
and $\Sigma NN$ states for the case
of pure $S$ wave configurations assuming that particle 1 is
the hyperon and particles 2 and 3 are the two nucleons. In order to
include arbitrary orbital angular momentum configurations we consider
the total angular momentum and total isospin $J$ and $I$ while
$\sigma _{1}$ ($\tau _{1})$ and $\sigma _{3}$ ($\tau _{3})$ stand for
the spin (isospin) of the hyperon and the nucleon respectively.
In addition, $\ell_i$, $s_i$,
$j_i$, $i_i$, $\lambda_i$, and $J_i$ are the orbital angular momentum,
spin, total angular momentum, and isospin of the pair $jk$ while
$\lambda_i$ is the orbital angular momentum between particle
$i$ and the pair $jk$ and $J_i$ is the result of coupling $\lambda_i$
and $\sigma_i$. If in Eqs. (10)$-$(14) of Ref. \cite{GFV07} 
we make the replacements

\begin{equation}
\{ns_2i_2\}\to \{n\ell_2s_2j_2i_2\lambda_2J_2\}\equiv\gamma_2,
\end{equation}

\begin{equation}
\{ms_3i_3\}\to \{m\ell_3s_3j_3i_3\lambda_3J_3\}\equiv\gamma_3,
\end{equation}

\begin{equation}
\{rs_1i_1\}\to \{r\ell_1s_1j_1i_1\lambda_1J_1\}\equiv\gamma_1,
\end{equation}
the three-body equations become

\begin{eqnarray}
T_{2;JI;\beta}^{\gamma_2}(q_2) & = & 
B_{2;JI;\beta}^{\gamma_2}(q_2)  
+\sum_{\gamma_3} \int_0^\infty dq_3\,\left[
(-1)^{1+\ell_2+\sigma _{1}+\sigma _{3}-s_{2}+\tau _{1}+\tau _{3}-i_{2}} 
A_{23;JI}^{\gamma_2\gamma_3}(q_2,q_3;E) \right.
\nonumber \\
& + &  \left. 2\sum_{\gamma_1} \int_0^\infty dq_1\,
A_{31;JI}^{\gamma_2\gamma_1}(q_2,q_1;E)
A_{13;JI}^{\gamma_1\gamma_3}(q_1,q_3;E) \right]
T_{2;JI;\beta}^{\gamma_3}(q_3),
\label{for20}
\end{eqnarray}
where $T_{2;SI;\beta}^{\gamma_2}(q_2)$ is a two-component vector

\begin{equation}
T_{2;JI;\beta}^{\gamma_2}(q_2) = \left( \matrix{
T_{2;JI;\Sigma\beta}^{\gamma_2}(q_2)  \cr
T_{2;JI;\Lambda\beta}^{\gamma_2}(q_2) \cr } \right),
\label{for2}
\end{equation}
while the kernel of Eq. (\ref{for20}) is a $2\times 2$ matrix defined by

\begin{equation}
A_{23;JI}^{\gamma_2\gamma_3}(q_2,q_3;E)=\left(\matrix{
A_{23;JI;\Sigma\Sigma}^{\gamma_2\gamma_3}(q_2,q_3;E)&
A_{23;JI;\Sigma\Lambda}^{\gamma_2\gamma_3}(q_2,q_3;E)\cr
A_{23;JI;\Lambda\Sigma}^{\gamma_2\gamma_3}(q_2,q_3;E)&
A_{23;JI;\Lambda\Lambda}^{\gamma_2\gamma_3}(q_2,q_3;E)\cr }\right),
\label{gl2}
\end{equation}

\begin{equation}
A_{31;JI}^{\gamma_2\gamma_1}(q_2,q_1;E)=\left(\matrix{
A_{31;JI;\Sigma N(\Sigma)}^{\gamma_2\gamma_1}(q_2,q_1;E)&
A_{31;JI;\Sigma N(\Lambda)}^{\gamma_2\gamma_1}(q_2,q_1;E)\cr
A_{31;JI;\Lambda N(\Sigma)}^{\gamma_2\gamma_1}(q_2,q_1;E)&
A_{31;JI;\Lambda N(\Lambda)}^{\gamma_2\gamma_1}(q_2,q_1;E)\cr}\right),
\label{gl3}
\end{equation}

\begin{equation}
A_{13;JI}^{\gamma_1\gamma_3}(q_1,q_3;E)  =\left(\matrix{
A_{13;JI;N\Sigma}^{\gamma_1\gamma_3}(q_1,q_3;E)  &
   0 \cr
   0&
A_{13;JI;N\Lambda}^{\gamma_1\gamma_3}(q_1,q_3;E)  \cr}\right) \, ,
\label{gl4}
\end{equation}
where

\begin{eqnarray}
A_{23;JI;\alpha\beta}^{\gamma_2\gamma_3}(q_{2},q_{3};E)
&=&\sum_{\ell_2^\prime r}
\tau_{2;\ell_2\ell_2^\prime s_{2}j_2i_{2};\alpha\beta}^{nr}
(E-q_{2}^{2}/2\nu _{2}){\frac{q_{3}^{2}}{2}}  \nonumber \\
&&\times \int_{-1}^{1}d{\rm cos}\theta \,{\frac{P_{r}(x^\prime_{2})
D_{23;JI;\beta}^{\rho_2^\prime\rho_3}(q_{2},q_{3},{\rm cos}\theta)
P_{m}(x_{3})}{%
E+\Delta E\delta_{\beta\Lambda}-p_{3}^{2}/2\mu _{3}-q_{3}^{2}/2\nu _{3}
+ i\epsilon}}; \,\,\,\,\,\,\,\, \alpha,\beta=\Sigma,\Lambda,
\label{gl5}
\end{eqnarray}

\begin{eqnarray}
A_{31;JI;\alpha N(\beta)}^{\gamma_2\gamma_1}(q_{2},q_{1};E)
&=&\sum_{\ell_2^\prime r}
\tau_{3;\ell_2\ell_2^\prime s_{2}j_2i_{2};\alpha\beta}^{nr}
(E-q_{2}^{2}/2\nu _{2})
{\frac{q_{1}^{2}}{2}}  \nonumber \\
&&\times \int_{-1}^{1}d{\rm cos}\theta \,{\frac{P_{r}(x^\prime_{3})
D_{31;JI;\beta}^{\rho_2^\prime\rho_1}(q_{2},q_{1},{\rm cos}\theta)
P_{m}(x_{1})}{%
E+\Delta E\delta_{\beta\Lambda}-p_{1}^{2}/2\mu _{1}-q_{1}^{2}/2\nu _{1}
+ i\epsilon}}; \,\,\,\,\,\,\,\, \alpha,\beta=\Sigma,\Lambda,
\label{gl6}
\end{eqnarray}

\begin{eqnarray}
A_{13;JI;N\beta}^{\gamma_1\gamma_3}(q_{1},q_{3};E)
&=&\sum_{\ell_1^\prime r}
\tau_{1;\ell_1\ell_1^\prime s_{1}j_{1}i_{1};NN}^{nr}
(E+\Delta E\delta_{\beta\Lambda}-q_{1}^{2}/2\nu _{1})
{\frac{q_{3}^{2}}{2}}  \nonumber \\
&&\times \int_{-1}^{1}d{\rm cos}\theta \,{\frac{P_{r}(x^\prime_{1})
D_{13;JI;\beta}^{\rho_1\prime\rho_3}(q_{1},q_{3},{\rm cos}\theta)
P_{m}(x_{3})}{%
E+\Delta E\delta_{\beta\Lambda}-p_{3}^{2}/2\mu _{3}-q_{3}^{2}/2\nu _{3}.  
+ i\epsilon}}; \,\,\,\,\,\,\,\, \beta=\Sigma,\Lambda,
\label{gl7}
\end{eqnarray}
where

\begin{equation}
\rho_i\equiv \{\ell_is_ij_ii_i\lambda_iJ_i\},
\end{equation}

\begin{equation}
\rho_i^\prime\equiv \{\ell_i^\prime s_ij_ii_i\lambda_iJ_i\},
\end{equation}
and $\eta_i$ and $\nu_i$ are the usual reduced masses
\begin{eqnarray}
\eta_i & = & {m_j m_k \over m_j + m_k} \, , \nonumber \\
\nu_i & = & {m_i(m_j+m_k) \over m_i+m_j+m_k}.
\label{redm}
\end{eqnarray}

In Eqs. (\ref{gl5})$-$(\ref{redm}) the isospin and mass of particle 1 
(the hyperon) is
determined by the subindex $\beta$. The
subindex $\alpha N(\beta)$ in 
Eq. (\ref{gl6}) indicates a transition $\alpha N \to \beta N$ with a
nucleon as spectator followed by a $NN \to NN$ transition with $\beta$ as
spectator. 
The angular momentum functions 
$D_{ij;JI;\beta}^{\rho_i\rho_j}(q_{i},q_{j},{\rm cos}\theta)$
are given by

\begin{eqnarray}
D_{ij;JI;\beta}^{\rho_i\rho_j}(q_{i},q_{j},{\rm cos}\theta) &=&
(-)^{i_j+\tau_j-I}\sqrt{(2i_i+1)(2i_j+1)} \, W(\tau_j\tau_kI\tau_i;i_ii_j)
\nonumber \\ && \times
\sqrt{(2j_i+1)(2j_j+1)(2J_i+1)(2J_j+1)}
\nonumber \\ && \times
\sum_{LS}(2L+1)(2S+1)
\left\{\matrix{
\ell_i & \lambda_i & L \cr
s_i & \sigma_i & S \cr
j_i & J_i & J \cr }\right\}
\left\{\matrix{
\ell_j & \lambda_j & L \cr
s_j & \sigma_j & S \cr
j_j & J_j & J \cr }\right\}
\nonumber \\ && \times
(-)^{s_j+\sigma_j-S}\sqrt{(2s_i+1)(2s_j+1)} \, W(\sigma_j\sigma_kS\sigma_i;s_is_j)
\nonumber \\ && \times
{1\over 2L+1}\sum_{M m_i m_j}C^{\ell_i \lambda_i L}_{m_i,M-m_i,M}
C^{\ell_j \lambda_j L}_{m_j,M-m_j,M}\Gamma_{\ell_i m_i}\Gamma_{\lambda_i
M-m_i}
\nonumber \\ && \times
\Gamma_{\ell_j m_j} 
\Gamma_{\lambda_j M-m_j}
{\rm cos}(-M\theta-m_i\theta_i
+m_j\theta_j),
\label{e8c5}
\end{eqnarray}
where $W$ is the Racah coefficient and $\Gamma_{\ell m}=0$ if $\ell -m$ 
is odd while
\begin{equation}
\Gamma_{\ell m}={(-)^{(\ell+m)/2}
\sqrt{(2\ell+1)(\ell+m)!(\ell-m)!}
\over 2^\ell((\ell+m)/2)!((\ell-m)/2)!},
\label{e10c5}
\end{equation}
if $\ell-m$ is even. The angles $\theta$, $\theta_i$, and $\theta_j$
are given by
\begin{equation}
{\rm cos}\theta={\vec q_i\cdot\vec q_j\over q_i q_j},
\end{equation}
\begin{equation}
{\rm cos}\theta_i={\vec q_i\cdot\vec p_i\over q_i p_i},
\end{equation}
\begin{equation}
{\rm cos}\theta_j={\vec q_j\cdot\vec p_j\over q_j p_j},
\end{equation}
with
\begin{eqnarray}
\vec p_i &=& - \vec q_j - {\eta_i\over m_k}\vec q_i \, , \nonumber \\
\vec p_j &=& \vec q_i + {\eta_j\over m_k} \vec q_j.
\label{e11c5}
\end{eqnarray}
$\tau_{i;\ell_i\ell_i^\prime s_ij_ii_i;\alpha\beta}^{nr}(e)$ 
are the coefficients of the 
expansion in terms of Legendre polynomials of the hyperon-nucleon
$t-$matrix $t_{i;\ell_i\ell_i^\prime s_ij_ii_i;\alpha\beta}(p_i,p_i^\prime;e)$ 
for the 
transition $\alpha N \to \beta N$, i.e.,

\begin{equation}
\tau_{i;\ell_i\ell_i^\prime s_ij_ii_i;\alpha \beta}^{nr}(e)=
{2n+1 \over 2}\,{2r+1 \over 2}
\int_{-1}^1 dx_i \int_{-1}^1
dx^\prime_i\, P_n(x_i)t_{i;\ell_i\ell_i^\prime s_ij_ii_i;\alpha \beta}
(p_i,p^\prime_i;e)
P_r(x^\prime_i) \, .
\label{for112}
\end{equation}

\noindent
The energy shift, $\Delta E$, is 
chosen such that at the $\beta d$ threshold the momentum
of the $\alpha d$ system has the correct value,
i.e., 

\begin{equation}
\Delta E={[(m_\beta+m_d)^2-(m_\alpha+m_d)^2]
[(m_\beta+m_d)^2-(m_\alpha-m_d)^2]\over
8\mu_{\alpha d}(m_\beta+m_d)^2},
\label{gl11}
\end{equation}
where $\mu_{\alpha d}$ is the $\alpha d$ reduced mass.

The inhomogeneous term of Eq. (\ref{for20}), 
$B_{2;JI;\beta}^{\gamma_2}(q_2)$ is a two-component vector

\begin{equation}
B_{2;JI;\beta}^{\gamma_2}(q_2) = \left( \matrix{
B_{2;JI;\Sigma\beta}^{\gamma_2}(q_2)  \cr
B_{2;JI;\Lambda\beta}^{\gamma_2}(q_2) \cr } \right),
\label{for3}
\end{equation}
where

\begin{eqnarray}
B_{2;JI;\alpha\beta}^{\gamma_2}(q_{2})
&=&\sum_{\ell_2^\prime r\rho_{10}}
\tau_{2;\ell_2\ell_2^\prime s_{2}j_2i_{2};\alpha\beta}^{nr}
(E_\beta^{th}-q_2^2/2\nu_2)
  \nonumber \\
&&\times P_{r}(x^\prime_{2})
D_{31;JI;\beta}^{\rho_2^\prime\rho_{10}}(q_{2},0,0)
\phi_{d;l_1}(q_2),
\label{for4}
\end{eqnarray}
and
\begin{equation}
\rho_{10}\equiv \{\ell_1,s_1=1,j_1=1,i_1=0,\lambda_1=0,J_1\},
\end{equation}
which corresponds to a hyperon-deuteron initial state,
$\phi_{d;\ell_1}(q_2)$ is the deuteron wave
function with orbital angular momentum $\ell_1$, 
$E_\beta^{th}$ is the energy of the $\beta d$ threshold,
$P_r(x_2^\prime)$ is a Legendre polynomial
of order $r$, and

\begin{equation}
x_2^\prime={{\eta_2\over m_3}q_2 - b\over {\eta_2\over m_3}q_2 + b}.
\label{for5}
\end{equation}

Finally, after solving the inhomogeneous set of equations (\ref{for20}),
the $\beta d$ scattering length is given by

\begin{equation}
A_{\beta d}=-\pi\mu_{\beta d}T_{\beta\beta},
\label{for6}
\end{equation}
with

\begin{equation}
T_{\beta\beta}=2\sum_{n\rho_{10}\rho_2}\int_0^\infty
q_2^2 dq_2\, \phi_{d;\ell_1}(q_2)
P_n(x_2^\prime) 
D_{13;JI;\beta}^{\rho_{10}\rho_2}(0,q_{2},0)
T_{2;JI;\beta\beta}^{\gamma_2}(q_2).
\label{for7}
\end{equation}
In the case of the $\Sigma NN$ system,
even for energies below the $\Sigma d$ threshold, 
one encounters the three-body singularities of the
$\Lambda NN$ system so that to solve the integral equations (\ref{for20})
one has to use the contour rotation method where the momenta are 
rotated into the complex plane $q_i\to q_i e^{-i\phi}$ since as pointed
out in Ref. \cite{Ter06} the results do not depend on the contour rotation
angle $\phi$. 

We give in Table \ref{t1} the two-body channels that contribute in the case of
the six three-body channels $(I,J)$ with $I=0,1,2$ and $J=1/2,3/2$. For the
parameter $b$ in Eqs. (\ref{for9}) and (\ref{for9p}) we found that 
$b=3$ fm$^{-1}$ leads to very stable results while for the expansion
(\ref{for11}) we took twelve Legendre polynomials, i.e., $0\le n\le 11$.

\section{Results}
\label{res}

In Ref. \cite{GFV07} we constructed 
different families of interacting potentials,
by introducing small variations of the mass of the
effective scalar exchange potentials, 
that allow us to study the dependence of the results on the
strength of the spin-singlet and spin-triplet 
hyperon-nucleon interactions. 
These potentials are characterized by
the $\Lambda N$ scattering lengths $a_{i,s}$
and they reproduce
the cross sections near threshold of the five
hyperon-nucleon processes for which data are 
available (see Ref. \cite{GFV07}).

\subsection{The $\Lambda NN$ system}

\begin{figure}[tbp]
\caption{Inverse of the $(I,J)=(0,1/2)$ 
and $(0,3/2)$ $\Lambda d$ scattering lengths
as a function of the $\Lambda N$ $a_{1/2,1}$ scattering length.}
\mbox{\epsfxsize=140mm\epsffile{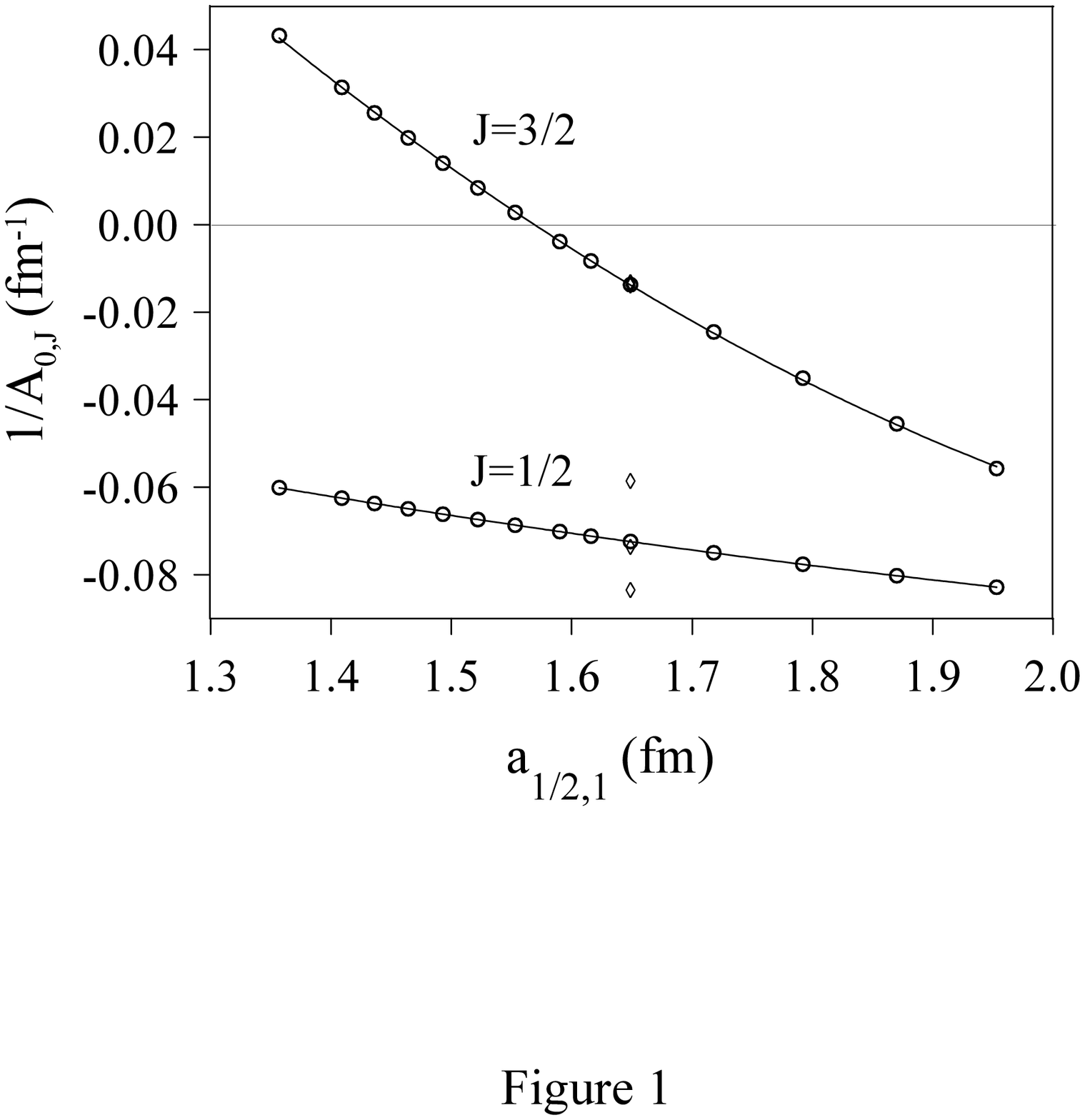}}
\label{fig1}
\end{figure}

The channels $(I,J)$ = (0,1/2) and (0,3/2) are the most attractive ones of
the $\Lambda NN$ system. In particular, the channel (0,1/2) has the only 
bound state of this system, the hypertriton. We give in Table \ref{t2}
the results of the models constructed in Ref. \cite{GFV07} for the
two $\Lambda d$ scattering lengths and the hypertriton binding energy.
We compare with the results, in parentheses, obtained in Ref.
\cite{GFV07} including only the three-body $S$ wave configurations.
As a consequence of considering the $D$ waves, the hypertriton binding energy
increases by about 50$-$60 keV \cite{Fuji}, 
while the $A_{0,1/2}$ scattering length
decreases by about 3$-$5 fm. The largest changes occur in the $A_{0,3/2}$
scattering length where both positive and negative values appeared which
means, in the case of the negative values, that a bound state is
generated in the $(I,J)=(0,3/2)$ channel. Since this channel depends
mainly on the spin-triplet hyperon-nucleon interaction and experimentally
there is no evidence whatsoever for the existence of a $(I,J)=(0,3/2)$ 
bound state
one can use the results of this channel to set limits on the value of
the hyperon-nucleon spin-triplet scattering length $a_{1/2,1}$. We plot
in Fig. \ref{fig1} the inverse of the two $\Lambda d$ scattering lengths as
a function of the spin-triplet $\Lambda N$ scattering length $a_{1/2,1}$.
As one can see, increasing
$a_{1/2,1}$ one increases the amount of attraction that is
present in the system since the
three-body channel $(I,J)=(0,3/2)$ becomes bound
if $a_{1/2,1} > 1.58$ fm.
Moreover, we found in Ref.~\cite{GFV07} that the fit 
of the hyperon-nucleon cross sections is worsened 
for those cases where the spin-triplet $\Lambda N$
scattering length is smaller than 1.41 fm,
so that we conclude that $1.41\le a_{1/2,1} \le 1.58$ fm. This range
of values is narrower than the one found in Ref.~\cite{GFV07}.

In order to show the dependence of these results on the spin-singlet
$\Lambda N$ scattering length $a_{1/2,0}$ we have also plotted 
in Fig.~\ref{fig1} the
results of the last three rows of Table~\ref{t2} where $a_{1/2,1}=1.65$ 
fm and $a_{1/2,0}$ = 2.31, 2.55, and 2.74 fm (they are denoted by 
{\it diamonds}). As one can see,
$1/A_{0,3/2}$ almost does not change although there is a large
sensitivity in $1/A_{0,1/2}$. In order to try to set some limits to the
hyperon-nucleon spin-singlet scattering length,
we have calculated in Table~\ref{t2p} the
hypertriton binding energy using for the hyperon-nucleon spin-triplet
scattering length the allowed values $1.41\le a_{1/2,1}\le 1.58$ fm
and using for the spin-singlet scattering length 
$2.33\le a_{1/2,0}\le 2.48$ fm which leads to results for the 
hypertriton binding energy within the experimental error bars
$B_{0,1/2}=0.13\pm 0.05$ MeV.

\begin{figure}[tbp]
\caption{Fredholm determinant for  the $\Lambda NN$
channels $(I,J) = (1,1/2)$ and $(1,3/2)$ for 
the model with $a_{1/2,0}=2.48$ fm and $a_{1/2,1}=1.41$ fm
and energies below the $\Lambda NN$ threshold.}
\mbox{\epsfxsize=140mm\epsffile{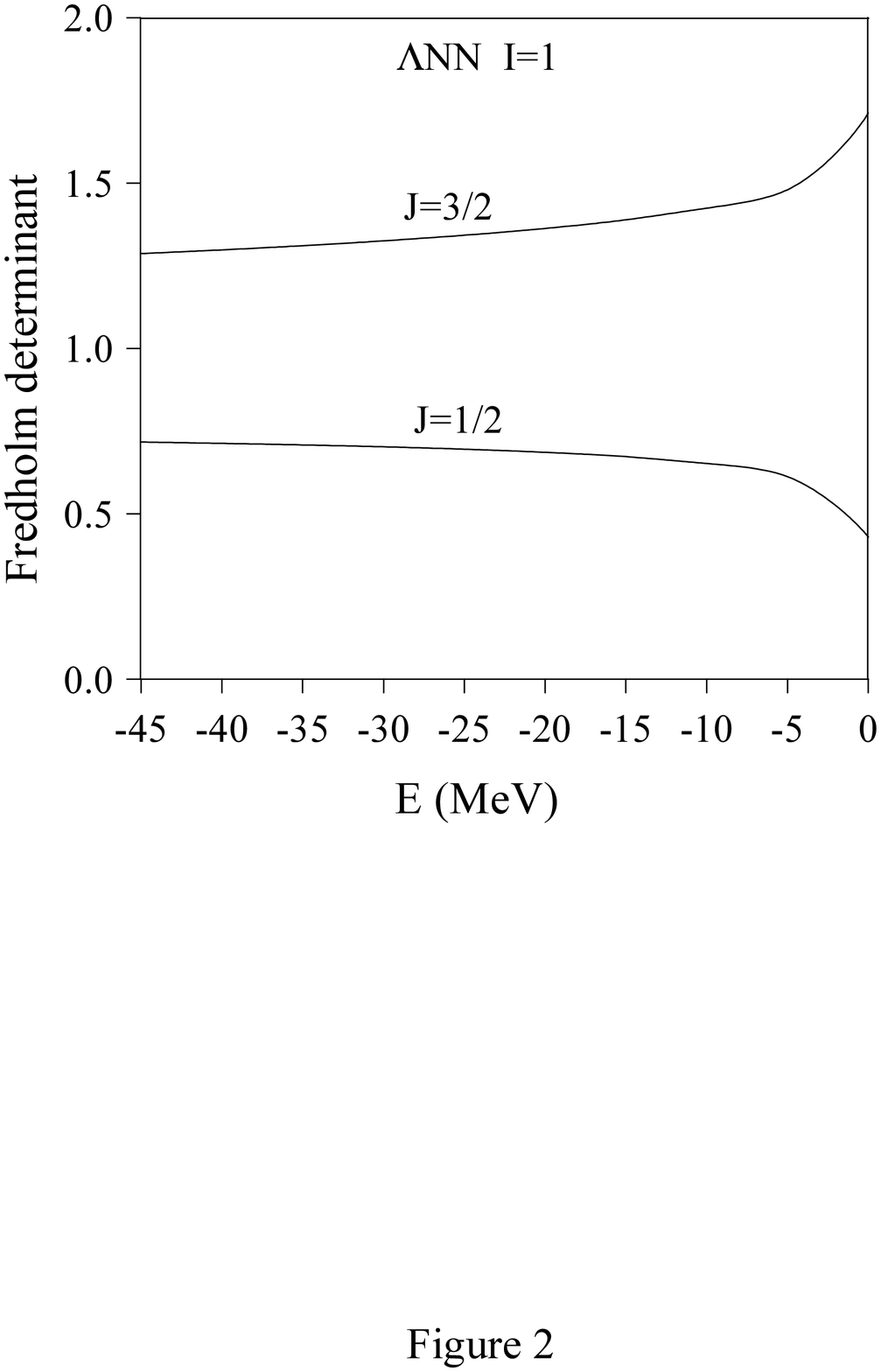}}
\label{fig2}
\end{figure}

With regard to the isospin 1 channels $(I,J)=(1,1/2)$ and (1,3/2), we show 
in Fig.~\ref{fig2} the Fredholm determinant of these channels for energies below
the $\Lambda NN$ threshold where one sees that the $(1,1/2)$ channel
is attractive but not enough to produce a bound state while the $(1,3/2)$
channel is repulsive. These results are very similar to the ones 
found in Ref.~\cite{GFV07}. 

\subsection{The $\Sigma NN$ system}

We show in Table \ref{t3} the $\Sigma d$ scattering lengths 
$A_{1,3/2}^\prime$ and $A_{1,1/2}^\prime$.
The $\Sigma d$ scattering lengths are 
complex since the inelastic $\Lambda NN$ channels are always open.
The scattering length $A_{1,3/2}^\prime$ depends mainly on the
spin-triplet hyperon-nucleon channels and both its real and
imaginary parts increase when the spin-triplet hyperon-nucleon
scattering length increases. The effect of 
the three-body $D$ waves is to lower 
the real part by about 20 \% and the imaginary part by about 10 \%. 
The scattering length $A_{1,1/2}^\prime$ shows large variations 
between the results with and without 
three-body $D$ waves but this is due, as
we will see next, to the fact that there is a pole very near threshold,
a situation quite similar to that of the $A_{0,3/2}$ $\Lambda d$
scattering length discussed in the previous subsection. 

\begin{figure}[tbp]
\caption{Real and imaginary parts of the $\Sigma d$ scattering
length $A_{1,1/2}^\prime$
as a function of the $\Lambda N$ $a_{1/2,1}$ scattering length.}
\mbox{\epsfxsize=140mm\epsffile{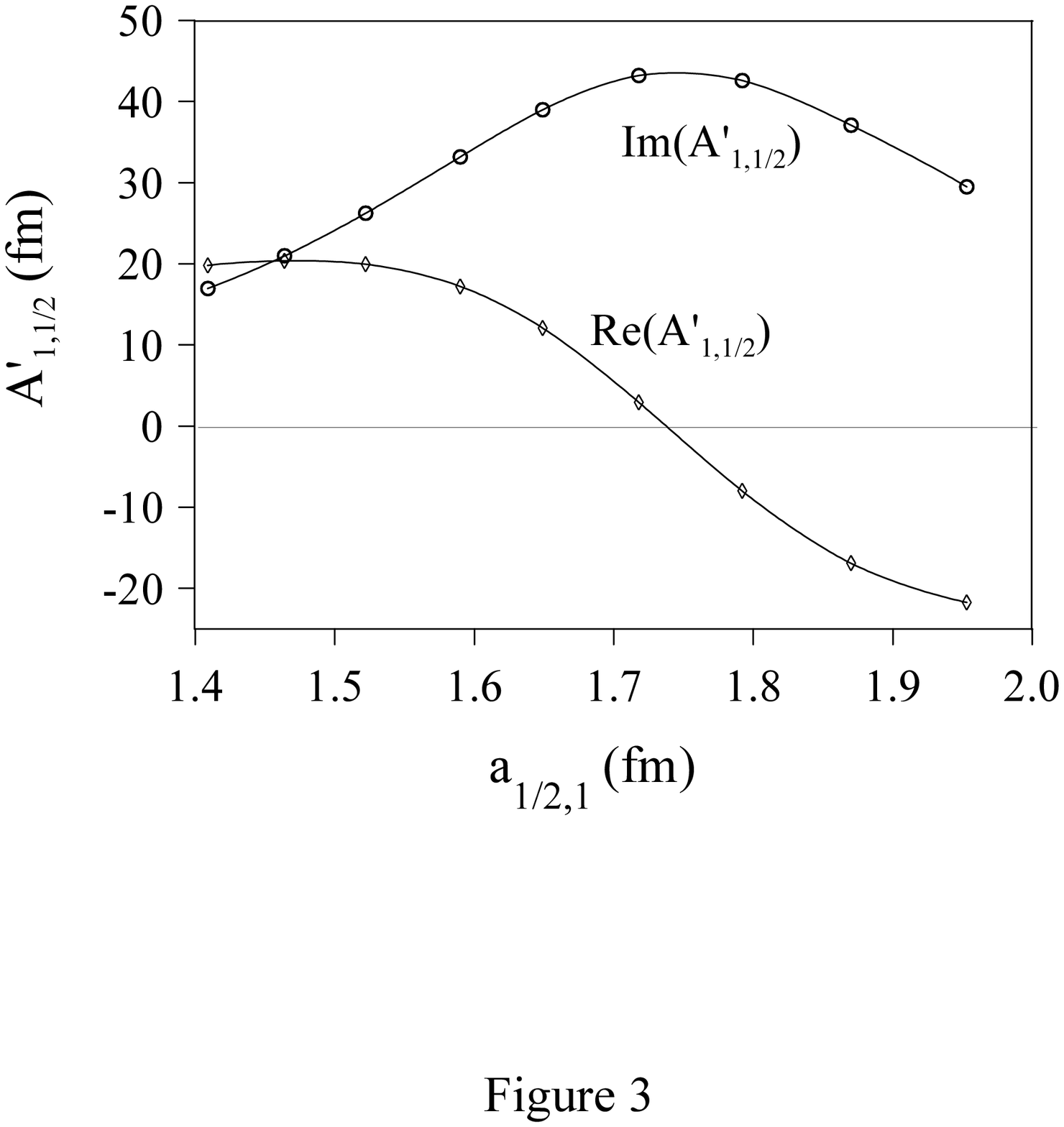}}
\label{fig3}
\end{figure}

\begin{figure}[tbp]
\caption{Diagram that gives the most important contribution to
the width of the $\Sigma d$ $(I,J)=(1,1/2)$ quasibound state.}
\mbox{\epsfxsize=140mm\epsffile{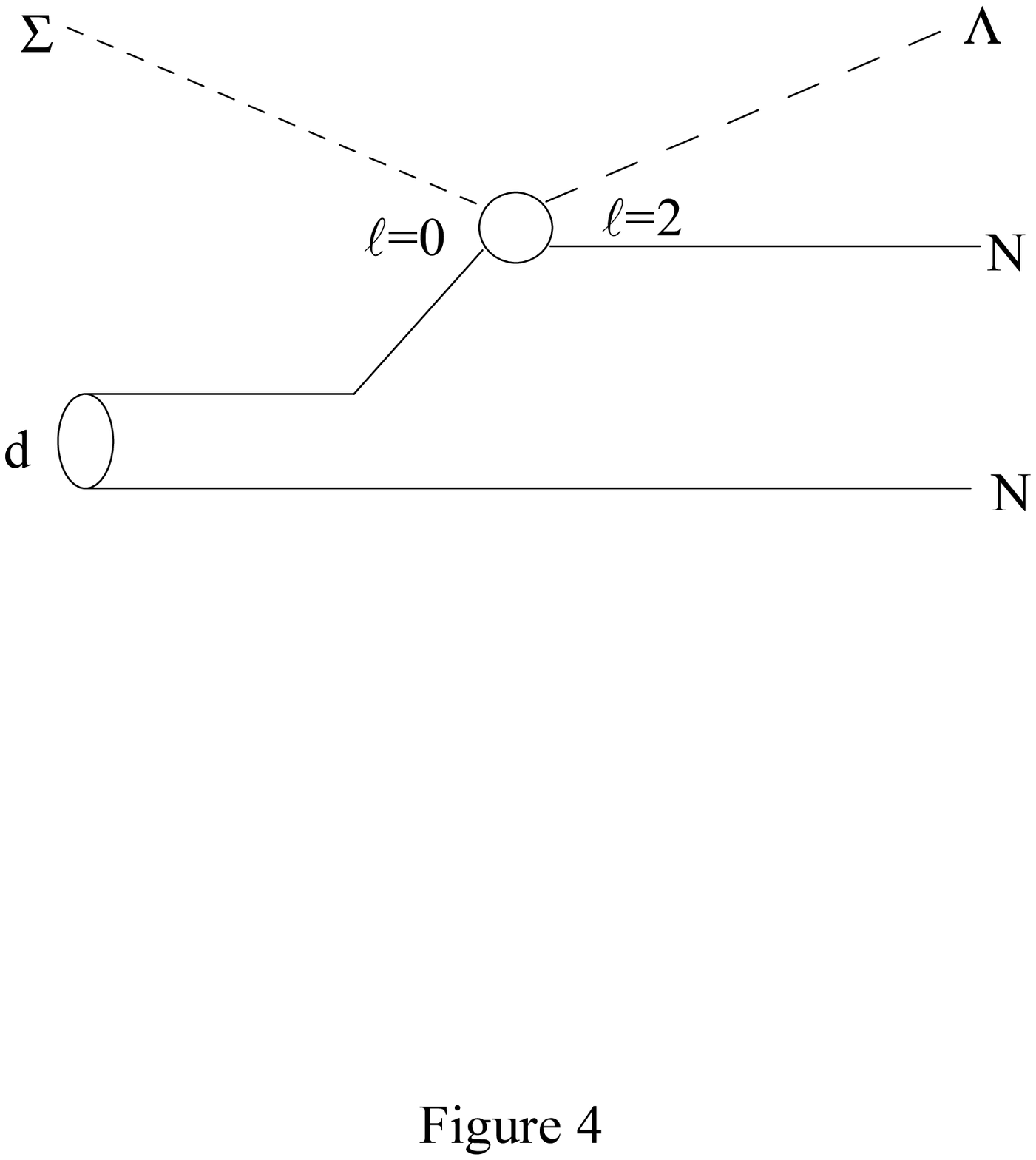}}
\label{fig4}
\end{figure}

We plot in Fig. \ref{fig3} the real and imaginary parts of the $\Sigma d$
scattering length $A_{1,1/2}^\prime$ as functions of the spin-triplet
$\Lambda N$ scattering length $a_{1/2,1}$, since by increasing 
$a_{1/2,1}$ one is increasing the amount of attraction that is
present in the three-body channel. As one can see, Re$(A_{1,1/2}^\prime)$
changes sign going from positive to negative while at the same time
Im$(A_{1,1/2}^\prime)$ has a maximum. These two features are the typical
ones that signal that the channel has a quasibound state \cite{DELOF}.
Since in the case of the $\Sigma NN$ system we are using the contour 
rotation method which opens large portions of the second Riemann sheet
we can search for the position of 
this pole in the complex plane which is given in the last
column of Table \ref{t3}. As one can see the position of the pole changes
very little with the model used to calculate it and it lies at around
2.8$\, -i\, $2.1 MeV. The diagram that gives the most important
contribution to the width of this state is the one drawn in Fig. \ref{fig4},
since the process $\Sigma N \to \Lambda N$ is dominated by the
transition $^3S_1$ $\to$ $^3D_1$. For example, at $p^\Sigma_{\rm LAB}=40$
MeV/c, the on-shell transition potential 
$V_{\Sigma\Lambda}(^3S_1$ $\to$ $^3D_1)=4.542 \, 10^{-2}$ fm$^2$,
while 
$V_{\Sigma\Lambda}(^3S_1$ $\to$ $^3S_1)=-1.008 \, 10^{-2}$ fm$^2$,
a factor four smaller.
The corresponding on-shell transition amplitudes are
$t_{\Sigma\Lambda}(^3S_1$ $\to$ $^3D_1)=8.520 \, 10^{-2}\, + \, {\rm i} \,
5.507 \, 10^{-2} $ fm$^2$,
and 
$t_{\Sigma\Lambda}(^3S_1$ $\to$ $^3S_1)=-1.061 \, 10^{-2}\, - \, {\rm i} \,
8.961 \, 10^{-3} $ fm$^2$, roughly a factor eight smaller.

\begin{figure}[tbp]
\caption{Fredholm determinant for (a) $J=1/2$ and (b) $J=3/2$ $\Sigma NN$
channels for the model with $a_{1/2,0}=2.48$ fm and $a_{1/2,1}=1.41$ fm.
The $\Sigma d$ continuum starts at $E=-2.225$ MeV, the
deuteron binding energy obtained within our model.}
\mbox{\epsfxsize=140mm\epsffile{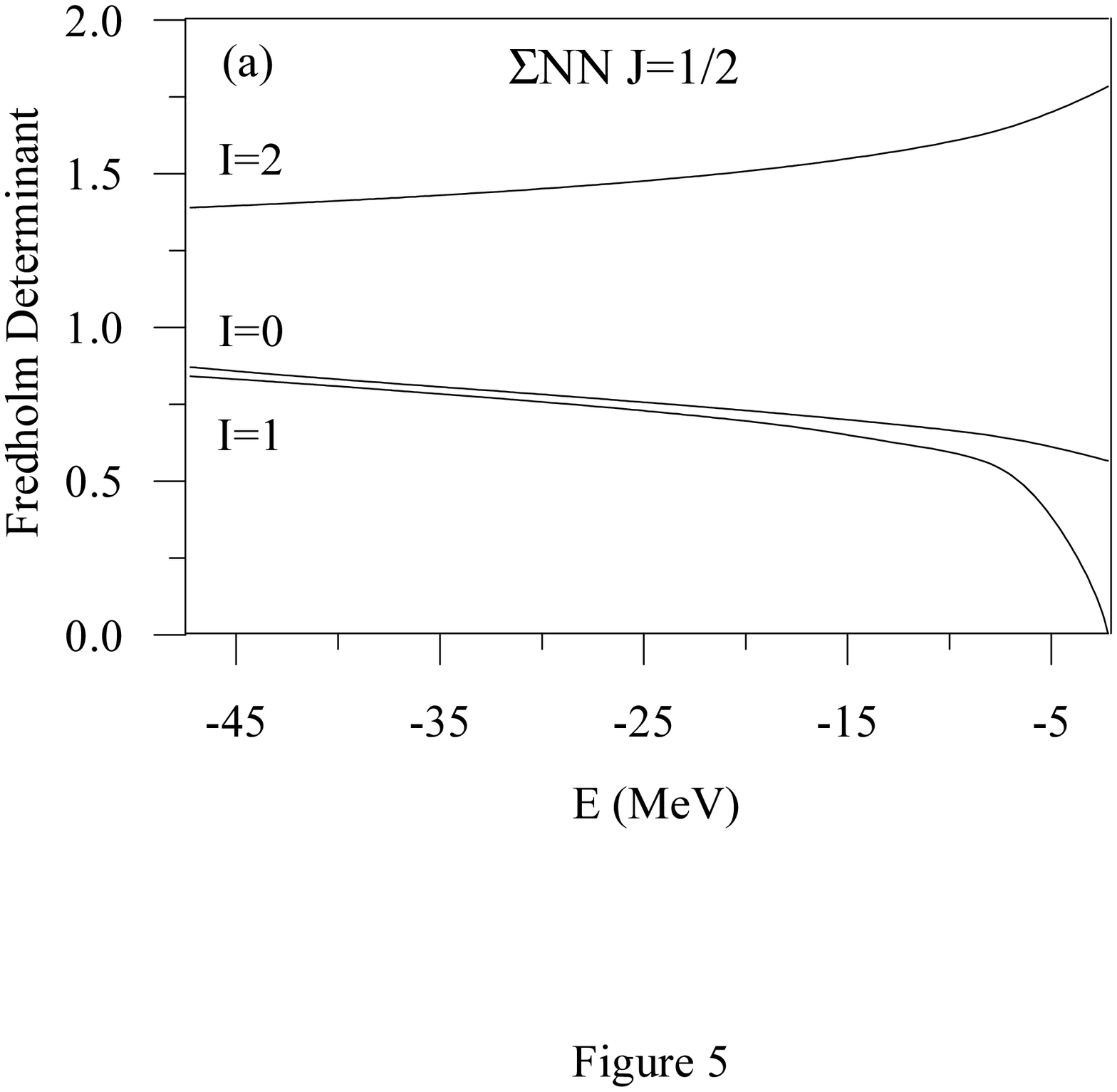}}
\newpage
\mbox{\epsfxsize=140mm\epsffile{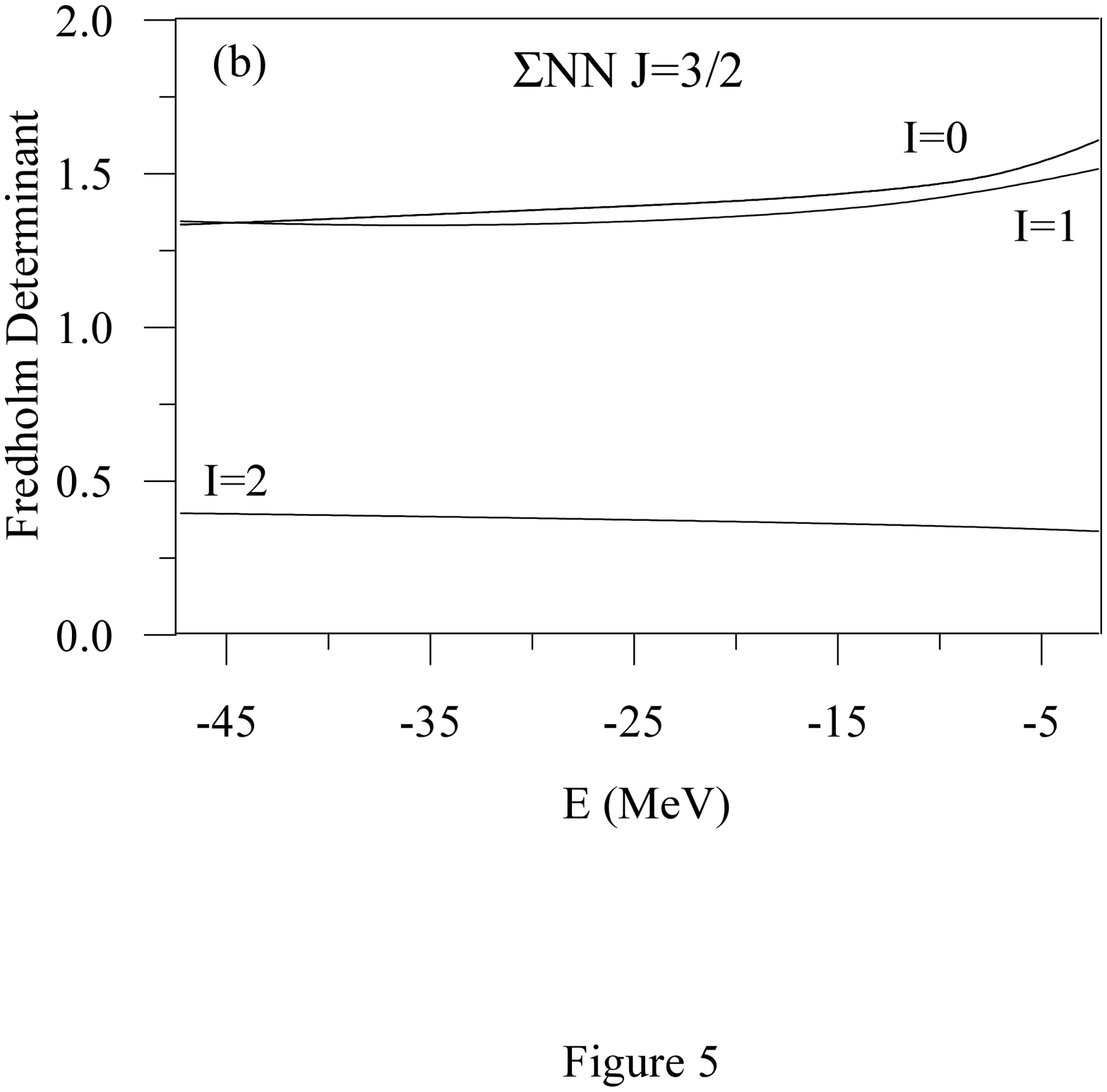}}
\label{fig5}
\end{figure}

We show in Fig. \ref{fig5} the real part of the Fredholm determinant of the six
$(I,J)$ $\Sigma NN$ channels that are possible for energies below the 
$\Sigma d$ threshold. The imaginary part of the Fredholm determinant
is small and uninteresting. As one can see the channel $(1,1/2)$
is the most attractive one since the Fredholm determinant is close to zero
at the $\Sigma d$ threshold, which as mentioned before,
indicates the presence of a quasibound state.
The next channel, in what to amount
of attraction is concerned, is the $(I,J)=(0,1/2)$. 
The ordering of the two attractive $\Sigma NN$ $J=1/2$
channels can be easily understood by looking at Table III
of Ref. \cite{GFV07}. All the attractive two-body channels
in the $NN$, $\Lambda N$, and  $\Sigma N$
subsystems contribute to the $(I,J)=(1,1/2)$ $\Sigma NN$ state (the $\Sigma N
$ channels $^{3}S_{1}(I=1/2)$ and $^{1}S_{0}(I=3/2)$ and the $^{3}S_{1}(I=0)$
$NN$ channel), while the $(I,J)=(0,1/2)$ state does not present contribution
from two of them, the $^{1}S_{0}(I=3/2)$ $\Sigma N$ and specially the
$^{3}S_{1}(I=0)$ $NN$  deuteron channel.

\section{Summary}

We have solved the Faddeev equations for the $\Lambda NN$ and
$\Sigma NN$ systems using the hyperon-nucleon and nucleon-nucleon
interactions derived from a chiral constituent quark model with
full inclusion of the $\Lambda \leftrightarrow \Sigma$ conversion
and taking into account all three-body configurations
with $S$ and $D$ wave components.

In the case of the $\Lambda NN$ system the inclusion of the 
three-body $D$ wave components increases the attraction, reducing
the upper limit of the $a_{1/2,1}$ $\Lambda N$ scattering length
if the $(I,J)=(0,3/2)$ $\Lambda NN$ bound state does not exist.
This state shows a somewhat larger sensitivity than the
hypertriton to the three-body $D$ waves. Our calculation
including the three-body $D$ wave configurations of 
all relevant observables of two- and three-baryon 
systems with strangeness $-1$, permits to constrain the 
$\Lambda N$ scattering lengths to:
$1.41 \le a_{1/2,1} \le 1.58$ fm and
$2.33 \le a_{1/2,0} \le 2.48$ fm.

In the case of the $\Sigma NN$ system there exists a narrow quasibound
state near threshold in the $(I,J)=(1,1/2)$ channel. The width of this
state, of the order of 2.1 MeV, comes mainly from the coupling to the
$\Lambda NN$ system in a $D$ wave three-body channel.

The actual interest in two- and three-baryon systems with strangeness
$-1$ \cite{EXP} makes worthwhile to pursue 
the experimental search of narrow peaks
near threshold related with the predictions of our model based on the
description of almost all known observables of the two- and three-baryons with
strangeness $-1$.

\acknowledgements
This work has been partially funded by Ministerio de Educaci\'{o}n y Ciencia
under Contract No. FPA2004-05616 and by COFAA-IPN (M\'{e}xico).

\begin{table}[tbp]
\caption{Two-body $\Sigma N$ channels with a nucleon as spectator 
$(\ell_\Sigma s_\Sigma j_\Sigma i_\Sigma\lambda_\Sigma J_\Sigma)_N$,
two-body $\Lambda N$ channels with a nucleon as spectator 
$(\ell_\Lambda s_\Lambda j_\Lambda i_\Lambda\lambda_\Lambda J_\Lambda)_N$,
two-body $NN$ channels with a $\Sigma$ as spectator 
$(\ell_N s_N j_N i_N \lambda_N J_N)_\Sigma$, and
two-body $NN$ channels with a $\Lambda$ as spectator 
$(\ell_N s_N j_N i_N \lambda_N J_N)_\Lambda$ that contribute to
a given $\Sigma NN - \Lambda NN$ state with total isospin $I$ and 
total angular momentum $J$.}
\label{t1}
\begin{tabular}{|cccccc|}
$I$ & $J$ & 
$(\ell_\Sigma s_\Sigma j_\Sigma i_\Sigma\lambda_\Sigma J_\Sigma)_N$ &
$(\ell_\Lambda s_\Lambda j_\Lambda i_\Lambda\lambda_\Lambda J_\Lambda)_N$
& $(\ell_N s_N j_N i_N \lambda_N J_N)_\Sigma$
& $(\ell_N s_N j_N i_N \lambda_N J_N)_\Lambda$ \\
\tableline 
0 & ${1\over 2}$ 
& (000${1\over 2}$0${1\over 2}$),(011${1\over 2}$0${1\over 2}$),  
& (000${1\over 2}$0${1\over 2}$),(011${1\over 2}$0${1\over 2}$),  
& (00010${1\over 2}$) & (01100${1\over 2}$),(21100${1\over 2}$), \\ 
& & (211${1\over 2}$0${1\over 2}$),(011${1\over 2}$2${3\over 2}$),  
& (211${1\over 2}$0${1\over 2}$),(011${1\over 2}$2${3\over 2}$),  
&  & (01102${3\over 2}$),(21102${3\over 2}$) \\ 
& & (211${1\over 2}$2${3\over 2}$)
&   (211${1\over 2}$2${3\over 2}$)
&  &  \\ \hline 
1 & ${1\over 2}$ 
& (000${1\over 2}$0${1\over 2}$),(011${1\over 2}$0${1\over 2}$),  
& (000${1\over 2}$0${1\over 2}$),(011${1\over 2}$0${1\over 2}$),  
& (00010${1\over 2}$),(01100${1\over 2}$), & (00010${1\over 2}$) \\ 
& & (211${1\over 2}$0${1\over 2}$),(011${1\over 2}$2${3\over 2}$),  
& (211${1\over 2}$0${1\over 2}$),(011${1\over 2}$2${3\over 2}$),  
&  (21100${1\over 2}$),(01102${3\over 2}$), & \\ 
& & (211${1\over 2}$2${3\over 2}$),(000${3\over 2}$0${1\over 2}$),  
& (211${1\over 2}$2${3\over 2}$)
& (21102${3\over 2}$) &  \\ 
& & (011${3\over 2}$0${1\over 2}$),(211${3\over 2}$0${1\over 2}$),  
& & & \\ 
& & (011${3\over 2}$2${3\over 2}$),(211${3\over 2}$2${3\over 2}$)  
& & & \\ \hline
2 & ${1\over 2}$ 
& (000${3\over 2}$0${1\over 2}$),(011${3\over 2}$0${1\over 2}$),  
& & (00010${1\over 2}$) &  \\ 
& & (211${3\over 2}$0${1\over 2}$),(011${3\over 2}$2${3\over 2}$),  
& &  & \\ 
& & (211${3\over 2}$2${3\over 2}$)
& &  &  \\ \hline
0 & ${3\over 2}$ 
& (000${1\over 2}$2${3\over 2}$),(011${1\over 2}$0${1\over 2}$),  
& (000${1\over 2}$2${3\over 2}$),(011${1\over 2}$0${1\over 2}$),  
& (00012${3\over 2}$) & (01100${1\over 2}$),(21100${1\over 2}$), \\ 
& & (211${1\over 2}$0${1\over 2}$),(011${1\over 2}$2${3\over 2}$),  
& (211${1\over 2}$0${1\over 2}$),(011${1\over 2}$2${3\over 2}$),  
&  & (01102${3\over 2}$),(01102${5\over 2}$), \\ 
& & (011${1\over 2}$2${5\over 2}$),(211${1\over 2}$2${3\over 2}$),  
& (011${1\over 2}$2${5\over 2}$),(211${1\over 2}$2${3\over 2}$),  
&  & (21102${3\over 2}$),(21102${5\over 2}$) \\ 
& & (211${1\over 2}$2${5\over 2}$)
&   (211${1\over 2}$2${5\over 2}$)
&  &  \\ \hline
1 & ${3\over 2}$ 
& (000${1\over 2}$2${3\over 2}$),(011${1\over 2}$0${1\over 2}$),  
& (000${1\over 2}$2${3\over 2}$),(011${1\over 2}$0${1\over 2}$),  
& (00012${3\over 2}$),(01100${1\over 2}$), & (00012${3\over 2}$) \\ 
& & (211${1\over 2}$0${1\over 2}$),(011${1\over 2}$2${3\over 2}$),  
& (211${1\over 2}$0${1\over 2}$),(011${1\over 2}$2${3\over 2}$),  
& (21100${1\over 2}$),(01102${3\over 2})$, & \\ 
& & (011${1\over 2}$2${5\over 2}$),(211${1\over 2}$2${3\over 2}$),  
& (011${1\over 2}$2${5\over 2}$),(211${1\over 2}$2${3\over 2}$),  
& (01102${5\over 2}$),(21102${3\over 2}$), & \\ 
& & (211${1\over 2}$2${5\over 2}$),(000${3\over 2}$2${3\over 2}$),  
& (211${1\over 2}$2${5\over 2}$)
& (21102${5\over 2}$) &  \\ 
& & (011${3\over 2}$0${1\over 2}$),(211${3\over 2}$0${1\over 2}$),  
& &  &  \\ 
& & (011${3\over 2}$2${3\over 2}$),(011${3\over 2}$2${5\over 2}$),  
& &  &  \\ 
& & (211${3\over 2}$2${3\over 2}$),(211${3\over 2}$2${5\over 2}$)  
& &  &  \\ \hline
2 & ${3\over 2}$ 
& (000${3\over 2}$2${3\over 2}$),(011${3\over 2}$0${1\over 2}$),  
& & (00012${3\over 2}$) &  \\ 
& & (211${3\over 2}$0${1\over 2}$),(011${3\over 2}$2${3\over 2}$),  
& &  &  \\ 
& & (011${3\over 2}$2${5\over 2}$),(211${3\over 2}$2${3\over 2}$),  
& &  &  \\ 
& & (211${3\over 2}$2${5\over 2}$)
& &  &  \\ 
\end{tabular}
\end{table}

\begin{table}[tbp]
\caption{$\Lambda d$ scattering lengths, $A_{0,3/2}$ and $A_{0,1/2}$ (in fm),
and hypertriton binding energy,
$B_{0,1/2}$ (in MeV), for several hyperon-nucleon interactions
characterized by $\Lambda N$ scattering lengths
$a_{1/2,0}$ and $a_{1/2,1}$ (in fm). We give in parentheses the results
obtained in Ref. \protect\cite{GFV07} including only three-body $S$ wave
configurations.}
\label{t2}
\begin{tabular}{|ccccc|}
$a_{1/2,0}$  &  $a_{1/2,1}$ &
 $A_{0,3/2}$ & $A_{0,1/2}$  & $B_{0,1/2}$ \\ 
\hline\hline
2.48 & 1.41 & 31.9 (66.3)        & $-$16.0 ($-$20.0) & 0.129 (0.089) \\
2.48 & 1.65 & $-$72.8 (198.2)    & $-$13.8 ($-$17.2) & 0.178 (0.124) \\
2.48 & 1.72 & $-$40.8 ($-$179.8) & $-$13.3 ($-$16.6) & 0.192 (0.134) \\
2.48 & 1.79 & $-$28.5 ($-$62.7)  & $-$12.9 ($-$16.0) & 0.207 (0.145) \\
2.48 & 1.87 & $-$22.0 ($-$38.2)  & $-$12.5 ($-$15.4) & 0.223 (0.156) \\
2.48 & 1.95 & $-$17.9 ($-$27.6)  & $-$12.1 ($-$14.9) & 0.239 (0.168) \\
2.31 & 1.65 & $-$76.0 (198.2)    & $-$17.1 ($-$22.4) & 0.113 (0.070) \\
2.55 & 1.65 & $-$73.6 (198.2)    & $-$13.6 ($-$16.8) & 0.185 (0.130) \\
2.74 & 1.65 & $-$72.1 (198.2)    & $-$12.0 ($-$14.4) & 0.244 (0.182) \\
\end{tabular}
\end{table}

\begin{table}[tbp]
\caption{Hypertriton binding energy
(in MeV) for several hyperon-nucleon interactions
characterized by $\Lambda N$ scattering lengths
$a_{1/2,0}$ and $a_{1/2,1}$ (in fm) which are within the
experimental error bars $B_{0,1/2}=0.130\pm 0.050$ MeV.}
\label{t2p}
\begin{tabular}{|c|cccc|}
& $a_{1/2,1}=1.41$  &  $a_{1/2,1}=1.46$ 
& $a_{1/2,1}=1.52$  &  $a_{1/2,1}=1.58$ \\
\hline
$a_{1/2,0}=2.33$ & 0.080 & 0.087   & 0.096 & 0.106 \\
$a_{1/2,0}=2.39$ & 0.094 & 0.102   & 0.112 & 0.122 \\
$a_{1/2,0}=2.48$ & 0.129 & 0.140   & 0.152 & 0.164 \\
\end{tabular}
\end{table}

\begin{table}[tbp]
\caption{$\Sigma d$ scattering lengths,
$A_{1,3/2}^\prime$ and $A_{1,1/2}^\prime$ (in fm),
and position of the quasibound state $B_{1,1/2}^\prime$
(in MeV) for several hyperon-nucleon interactions
characterized by $\Lambda N$ scattering lengths
$a_{1/2,0}$ and $a_{1/2,1}$ (in fm). We give in parentheses the results
obtained in Ref. \protect\cite{GFV07} with only three-body $S$ waves.}
\label{t3}
\begin{tabular}{|ccccc|}
$a_{1/2,0}$  &  $a_{1/2,1}$ &
 $A_{1,3/2}^\prime$ & $A_{1,1/2}^\prime$ & $B_{1,1/2}^\prime$ \\ 
\hline\hline
2.48 & 1.41 & 0.14$+\, i\, $0.24 (0.20$\, +i\, $0.26) & 19.82$+\, i\, $16.94 (19.28$+\, i\, $25.37)       &2.92$-\, i\, $2.17 \\
2.48 & 1.65 & 0.28$+\, i\, $0.27 (0.36$\, +i\, $0.29) & 12.08$+\, i\, $38.98 ($-$1.55$+\, i\, $42.31)     &2.84$-\, i\, $2.14 \\
2.48 & 1.72 & 0.32$+\, i\, $0.28 (0.40$\, +i\, $0.30) & 2.92$+\, i\, $43.20 ($-$10.47$+\, i\, $40.25)     &2.82$-\, i\, $2.11 \\
2.48 & 1.79 & 0.36$+\, i\, $0.29 (0.44$\, +i\, $0.31) &$-$8.00$+\, i\, $42.58 ($-$17.33$+\, i\, $35.01)   &2.79$-\, i\, $2.10 \\
2.48 & 1.87 & 0.40$+\, i\, $0.30 (0.49$\, +i\, $0.33) &$-$16.90$+\, i\, $37.08 ($-$21.16$+\, i\, $28.54)  &2.77$-\, i\, $2.09 \\
2.48 & 1.95 & 0.45$+\, i\, $0.31 (0.54$\, +i\, $0.34) &$-$21.73$+\, i\, $29.48 ($-$22.44$+\, i\, $22.44)  &2.75$-\, i\, $2.08 \\
2.31 & 1.65 & 0.28$+\, i\, $0.27 (0.36$\, +i\, $0.29) & 19.01$+\, i\, $23.21 (14.95$+\, i\, $31.61)       &2.88$-\, i\, $2.14\\
2.55 & 1.65 & 0.28$+\, i\, $0.27 (0.36$\, +i\, $0.29) & $-$12.81$+\, i\, $43.49 ($-$21.04$+\, i\, $33.19) &2.79$-\, i\, $2.11\\
2.74 & 1.65 & 0.28$+\, i\, $0.27 (0.36$\, +i\, $0.29) & $-$26.01$+\, i\, $17.95 ($-$23.29$+\, i\, $13.32) &2.73$-\, i\, $2.09\\
\end{tabular}
\end{table}

\newpage

\newpage

\end{document}